\PassOptionsToPackage{unicode}{hyperref}
\PassOptionsToPackage{hyphens}{url}
\documentclass[12pt]{article}
\usepackage{amsmath,amssymb}
\usepackage{iftex}
\ifPDFTeX
  \usepackage[T1]{fontenc}
  \usepackage[utf8]{inputenc}
  \usepackage{textcomp} 
\else 
  \usepackage{unicode-math} 
  \defaultfontfeatures{Scale=MatchLowercase}
  \defaultfontfeatures[\rmfamily]{Ligatures=TeX,Scale=1}
\fi
\usepackage{lmodern}
\usepackage[margin=1in]{geometry}
\ifPDFTeX\else
\fi
\IfFileExists{upquote.sty}{\usepackage{upquote}}{}
\IfFileExists{microtype.sty}{
  \usepackage[]{microtype}
  \UseMicrotypeSet[protrusion]{basicmath} 
}{}
\makeatletter
\@ifundefined{KOMAClassName}{
  \IfFileExists{parskip.sty}{%
    \usepackage{parskip}
  }{
    \setlength{\parindent}{0pt}
    \setlength{\parskip}{6pt plus 2pt minus 1pt}}
}{
  \KOMAoptions{parskip=half}}
\makeatother
\usepackage{xcolor}
\usepackage[numbers,sort]{natbib} 
\ifLuaTeX
  \usepackage{selnolig}  
\fi
\IfFileExists{bookmark.sty}{\usepackage{bookmark}}{\usepackage{hyperref}}
\IfFileExists{xurl.sty}{\usepackage{xurl}}{} 
\hypersetup{
  pdftitle={Intelligent Interaction Techniques -- Proposal},
  hidelinks,
  pdfcreator={LaTeX via pandoc}}

\title{Intelligent Interaction Techniques (IIxT) -- Proposal}
\author{Brad A. Myers \\
Human-Computer Interaction Institute \\
School of Computer Science \\
Carnegie Mellon University \\
Pittsburgh, PA, USA \\
\texttt{bam@cs.cmu.edu}}

\date{}

\begin{document}
\maketitle
\begin{center}
\small \textcopyright\ 2026 Brad A. Myers
\end{center}
\bigskip 

\begin{abstract}
Interaction techniques (IxTs) are the low-level, reusable components out of which user interfaces are designed, including menus, scroll bars, text input fields, and also copy-paste, text-entry, and selecting objects. The IxTs for graphical user interfaces (GUIs) were well established in the 1980s, with relatively minor additions and tweaks for smartphones in the 2000s. Most of today's AI user interfaces involve a chat window, which is an excellent interaction for some tasks, but is generally considered separate from the GUI IxTs. I argue for making the IxTs \textit{themselves} more intelligent, so users can freely mix modalities, even within the same interaction. This will require research into new IxTs, and also into the infrastructure that will enable these intelligent IxTs (IIxTs) to be built. There are also significant security, privacy and economic implications to this vision.
\end{abstract}

\section*{Author Keywords}

Interaction Techniques (IxT); Intelligent Interaction Technique (IIxT), Widgets; Intelligent User Interfaces; Smart widgets.

%

\hypertarget{introduction}{%
\section{Introduction}\label{introduction}}

An interaction technique (IxT) is formally defined as starting when the user performs an action that causes an product to respond, and includes the direct feedback from the product to the user \cite[section 1.3.8]{Myers2024}. Interaction techniques are generally reusable across various applications. Examples of interaction techniques include on-screen widgets like menus, scrollbars and buttons, but also the low-level ways of interacting with on-screen objects, such as swiping gestures to scroll, copy-and-paste, drag-and-drop, text entry and editing, graphical selection and editing, etc. An important distinction is that \emph{applications}, like Microsoft Word, GMail, Facebook, or Yelp, are \emph{composed} out of (many) interaction techniques, and are not themselves considered an IxT. Most of the interaction techniques used in regular user interfaces, often called graphical user interfaces (GUIs), were developed in the 1970s and 1980s, and were updated slightly with smartphones in the 2000s. My book \cite{Myers2024} has a comprehensive discussion of this topic.

Interaction techniques are often collected in a library to make them easier for others to reuse. For example, low-level widgets are often provided by a \emph{toolkit} code library, and support for other IxTs like copy-paste and undo are provided by \emph{frameworks} \cite{Myers1995}.

What counts as making an IxT \emph{intelligent}? This is defined (somewhat vaguely) as an interaction technique that uses Artificial Intelligence (AI) to make it work \cite[section 18.2]{Myers2024}. One defining feature of all intelligent user interfaces is that previously deterministic IxTs will become probabilistic and therefore may be incorrect.

Intelligent user interface (IUI) or AI interaction research is often concerned with the \emph{whole} user interface and how it behaves, usually composed of conventional IxTs. For example, today's successful AI applications, such as Claude, Firefly in Adobe products, CoPilot in Microsoft products, Figma Design Agent \& Figma Make, etc., are mostly chat interfaces (which use a conventional text entry IxT) \emph{along side} conventional GUIs, rather than having the AI \emph{integrated} into the interaction techniques themselves.

There has always been at least some research on intelligent interaction techniques (IIxTs), often published at UIST and CHI, which has been inspirational (e.g., \cite{Bier1986,Stylos2004,Wobbrock2003,Yang2020,Zindulka2025}). However, commercial IxTs have used older AI techniques which were not particularly successful, such as the Microsoft Office XP ``personalized menus'' from 2001 which tried to be ``smart'' about which items to display.

I argue that with today's much better AI, including Large Language Models (LLMs), Multimodal LLMs (MLLMs), Generative AI (GenAI), etc., it is time to revisit how to create reusable, composable low-level interaction techniques that are intelligent and useful, especially for graphical user interfaces.

\hypertarget{why-guis-are-still-relevant}{%
\section{Why GUIs are Still Relevant}\label{why-guis-are-still-relevant}}

Some have argued that we will not use conventional graphical user interfaces, desktops or smartphone screens or their interaction techniques once natural language and speech interfaces get better. I disagree. There are definitely some tasks that are easier when performed using natural language, but there are others than are not. Many other writers have also made this argument (e.g., \cite{Gao2024,Shneiderman1997}), so I will just list a few points here. In later sections, I discuss how to augment IxTs to preserve their advantages while making them more intelligent.

Ben Shneiderman's original list of properties elucidates situations when \emph{direct manipulation} interfaces are important and useful: when it is feasible to have ``\textbf{continuous representation of the object of interest''} with ``\textbf{rapid, incremental, reversible operations whose impact on the object of interest is immediately visible}'' \cite{Shneiderman1983}. When GUIs were being invented in the 1970s and 1980s, and similarly at the beginning of the smartphone era in the 2000s, users had a small number of files and applications and their icons, so arranging or listing them on the screen was feasible. Today, even when we have hundreds to millions of items on our devices, there are times when it is still faster to list and tap on items, and other times when an intelligent search is preferred or necessary. For example, I have about 30 apps on the front screens of my smartphone that I regularly want to quickly tap on, but for most of the others, I switch modalities and type the name to search. I strongly prefer to find files by location in the folder tree, whereas I almost only find email messages by using textual search. And when we find the objects of interest, we usually want to see them in a list, so we can perform operations easily and quickly -- I want to hit the ``Delete'' keyboard key or the trashcan icon for most of my emails.

Another advantage of GUIs is that sometimes users do not know what can be done or do not know what things are called, so expecting them to express it, even in natural language, is problematic, compared to being able to browse a menu of options. The menu also helps teach users the correct vocabulary for the current task.

For understanding data, the visualization field \cite{Card1999} has shown that often understanding only comes from providing a visual representation that the user can view and interact with, rather than trying to describe the data in words.

Similarly, users may prefer to visually browse to find the items of interest if the distinguishing features are visual, rather than having to describe them using natural language. Producing (speaking or typing) and consuming (hearing or reading) natural language is still a linear, sequential process, whereas visual search and understanding is parallel, and can sometimes be performed in the background when the user's cognition is focused on a foreground task.

\hypertarget{historical-examples-of-intelligent-ixts}{%
\section{Historical Examples of Intelligent IxTs}\label{historical-examples-of-intelligent-ixts}}

As mentioned in the introduction, there is too much research on intelligent user interfaces in general to summarize here (see the IUI conference series, CHI, UIST and a few summaries \cite[chapter 18]{Myers2024} \cite{Gao2024,Volkel2020}). There are even too many intelligent IxTs to summarize here, but I present a few examples to help clarify what I mean by reusable interaction techniques which are intelligent.

The Put-That-There system \cite{Bolt1980} from the 1970s demonstrated an important early example of how speech recognition could enhance pointing, including when there were \emph{multiple} references in the same utterance (hence, ``put \emph{that} \emph{there}.'' Some IUI systems today allow the AI to access the selected object(s), but I don't know of any that support \emph{temporal} information in the selections and speech recognition to be able to support knowing which object goes with which part of the utterance.

The use of AI in text editing techniques dates back to at least 1966 with Warren Teitelman's \emph{Do What I Mean (DWIM)} spelling-correction system \cite{Teitelman1966}. The auto-correct, auto-complete and auto-fill capabilities of text editing have continually improved over time \cite[section 8.4]{Myers2024}, and now LLMs enable much longer completions (up to the whole document or code!), and more sophisticated editing \cite{Zindulka2025}, and are available in many places where users need to enter text.

A different kind of IIxT is highlighted by Adobe Photoshop and related products, where selection and operations on parts of pictures are augmented by image recognition. For example, one can select and extract just a face or a person from a photograph, intelligently differentiating their hair and body from the background. I classify this as an IxT since it can be available across any application that requires image selection or editing.

Finally, the areas of virtual reality (VR), augmented reality (AR), and mixed reality (MR), collectively called extended reality (XR), have long needed new interaction techniques. The user still needs to identify and operate on the objects in the 3D scenes, but conventional 2D GUI widgets do not work well. AI in the form of image, gesture and speech recognition, is often used in these environments to improve the accuracy and success of the UIs.

\hypertarget{proposed-examples-of-iixts}{%
\section{Proposed Examples of IIxTs}\label{proposed-examples-of-iixts}}

Virtually all interaction techniques \cite{Myers2024} could potentially be improved by being more intelligent. While some areas (like text entry) have received a lot of attention, others have not yet been investigated. Rather than trying to be comprehensive, this section presents some examples.

\hypertarget{intelligent-referencing}{%
\subsection{Intelligent Referencing}\label{intelligent-referencing}}

Indicating the object of interest, also called \textbf{pointing} \cite[chapter 4]{Myers2024} or \textbf{selecting} \cite[sections 8.6.3 and 10.5]{Myers2024}, has always been fundamental to user interfaces for all technology. Early research showed that the ``mouse'' pointing device \cite{English1967} was faster and more accurate than alternatives \cite{Card1978}, but current systems use less accurate pointing devices such as finger touches. Furthermore, users are now tasked with selecting among significantly more objects (e.g., among tens of thousands of email messages or photographs) and of more complex objects (to identify which part of the 3D avatar should be selected). There are many ways that this could be more intelligent.

Imagine if the user could simultaneously \emph{describe} the desired selection while pointing. For example, performing a low-resolution indication of an area, and expressing the details in natural language (``at the left of that character'', ``right-most point of that curve''). Besides just natural language clarifying the graphical properties, with today's AI, we can provide real ``semantic snapping'' \cite{Hudson1990} that takes into account the \emph{meaning} of the desired object (``the iris of the left eye'', ``just the name of the restaurant'', ``the red cars''). There could be easier ways to have \emph{multiple} selections (``add all of these'', ``select all the phone numbers in this paragraph''). These examples also suggest how users might \emph{fix} the selection set through a combination of direct manipulation (shift-touch to toggle items) and natural language (``undo the last one''). This might involve more intelligent checking for outliers to help users avoid selecting undesired objects \cite{Miller2002}. The selections could be saved, to help users reestablish their context (``take me to where I left off yesterday'')\footnote{Thanks to Haiyi Zhu for contributing this idea.}.

The pointing infrastructure and the natural language both need to do a better job of recording the temporal aspects, so users can refer to multiple items in a single utterance (``put that there''), but also so they can refer to past actions (``deselect the first one'').

Finally, we need better ways to \emph{refer} to items without necessarily operating on them, so users can tell the AI \emph{about} some items. Sometimes, systems must have different modes, like the Run versus Test modes in Interactive Builder programs \cite{Myers1995} like Microsoft Visual Basic or programming-by-example systems \cite{Myers1993}. We developed a ``wiggling'' interaction technique as one way to reference objects at a location without triggering the hovering or clicking actions on them \cite{Liu2022}.

\textbf{Searching} using text has always been an important way for users to find the items of interest. This has become the predominant method for many people to find files, email messages, and apps on smartphones, as the number of items has gotten too large to memorize or search visually. Intelligent searching using text is already becoming available, but this can become even more intelligent by enabling searching with other properties beyond the text of objects, such as the semantic properties mentioned above. Graphical search was researched by a few systems \cite{Kurlander1988,Myers1998b}, but was too hard to provide, which AI may make easier. Similarly, \emph{associative search} \cite{Chau2008}, where users can find items using meta properties like their creation time, location or other items they are related to, can also be provided.

\hypertarget{other-intelligent-widgets}{%
\subsection{Other Intelligent Widgets}\label{other-intelligent-widgets}}

Conventional widgets could also be made smarter. Smart \textbf{menus} were unsuccessful in Windows XP in 2001, in part because they did a bad job of predicting what the user might want, whereas today, such approaches can be much more successful. More interesting is integrating menus with semantic search, so users can express what they want, and the system would show users the menu options, keyboard shortcuts and icons for that, so users learn the faster methods to perform that action in the future. Intelligent \textbf{customization} of the menus and other commands could make it much easier for users to personalize their UIs. For example, previous versions of Microsoft Toolbars provided extensive customization, but this was removed with the introduction of the ribbon in Office 2007 \cite[section 9.5]{Myers2024}, because users messed it up too easily --- an intelligent mechanism might improve both the personalization and usability, so users could issue commands more efficiently.

Research has shown that for certain tasks, users waste a lot of time \textbf{scrolling} to find what they want to look at, and especially scrolling \emph{back} to where they were \cite{Ko2006}. An intelligent scrolling mechanism would combine scrolling manually with jumping to specific places in the document (like based on a table of contents), setting bookmarks in the documents, and with a mechanism for returning (undoing the scroll) \cite{Appert2012,Myers1998b}. It could also help coordinate scrolling of \emph{multiple} documents, for example, for comparing the related content across web pages or documents. Combining intelligent search and scrolling mechanisms might result in a generalized \textbf{intelligent navigation} mechanism.

Providing \textbf{undo} of interaction techniques could be generalized beyond scrolling -- I often want to undo moving or closing a window, undo a change to the selection, or of other actions which are currently not undoable. We have also lost the ability to undo many operations in web and mobile applications that we used to be able to undo on desktops, which should be restored intelligently. Techniques for \textbf{selective undo} have been demonstrated by research and a (very) few commercial systems, where the user can specify to undo only \emph{some} of the previous actions (e.g., ``Undo just the formatting changes, but keep the text edits'', ``Undo just the changes to this paragraph, but keep the changes elsewhere that I made afterwards'') \cite{Berlage1994,Myers1998b,Myers2015,Yoon2015}, but they have always been hampered by clunky UIs. An intelligent mechanism would allow users to simply express which operations to keep and undo.

Having an intelligent agent always available (like Apple Siri, Google Ask Gemini, or Microsoft Copilot), but which was aware of the interaction techniques and operations, could provide ubiquitous undo and \textbf{help} everywhere, and would provide a locus where there might be an intelligent \textbf{slider} for controlling how much autonomy is desired.\footnote{Thanks to Dan Saffer for contributing this idea.}

\hypertarget{intelligent-operations-across-applications}{%
\subsection{Intelligent Operations Across Applications}\label{intelligent-operations-across-applications}}

Today's cross-application interaction techniques are limited to copy-and-paste and drag-and-drop, which could be made more intelligent, and others could be added.

Intelligent selection could be the first step of intelligent \textbf{copy-and-paste}. In 2004, we investigated having the paste operation take into account the structure of the data on the clipboard, and allow it to be pasted into multiple places, like putting the parts of an address into different web form fields \cite{Stylos2004}, but with today's AI, many transformations of the data could be supported, including allowing users to express what is desired. For example, ``paste as bullet points'', ``paste as bibtex citations'', or ``paste translated into English''\footnote{Thanks to Motahhare Eslami for some of these examples.}: These could also be extended to support \textbf{drag-and-drop} as well.

Other operations \textbf{across applications} should be supported. Agents today have some limited abilities to remote control and collect information from some applications on the web, and to a lesser extent on the desktop and smartphones, but this should be expanded. One can easily say in natural language ``Group my flight, hotel and restaurant reservations together,'' but this is still difficult to perform if they are from independent applications.

\hypertarget{intelligent-accessibility}{%
\subsection{Intelligent Accessibility}\label{intelligent-accessibility}}

The hooks that would allow intelligent interaction techniques would be a boon for accessibility. Most low-level IxTs on major platforms today have lots of accessibility hooks and settings, but not typically IxTs for the web (like menus on web pages). Providing a natural-language interface that is available at a low-level everywhere would also be a tremendous advantage for vision-impaired and physically disabled individuals who can type or provide speech input.

\hypertarget{architectural-considerations-for-iixts}{%
\section{Architectural considerations for IIxTs}\label{architectural-considerations-for-iixts}}

Most interaction techniques of the types discussed here are provided in libraries and frameworks (with the exception of many kinds of menus in web and smartphone applications). Therefore, some of the proposed ideas could be realized entirely within those layers. However, most of the interesting forms of intelligence discussed above require knowledge about the applications themselves, and of the user. This imposes significant barriers from \textbf{architectural, security and privacy} perspectives. For example, I have been told that the reason that the text entry keyboard on Apple smartphones does not provide better proposed completions is that it is \emph{not} allowed to know much about the application for privacy reasons.

The IIxTs described here would likely be developed by the platform vendors who now provide the toolkits (Microsoft, Apple, Google). This means that they will be able to devote significant resources to their development and testing. However, IIxTs have even higher requirements for security and privacy than intelligent apps, since users will typically not be able to opt out of using them. Further, this might mean even more \textbf{consolidation and power} for the platform vendors.

To provide the proposed cross-application intelligent interactions discussed above, the agent will need access to information from those applications, which today is not available. APIs or screen scraping can provide to Siri or Claude some limited information and control of the application, but full access would require a new software architecture for applications besides the siloed apps of today. Researchers have called for this for many years (see my ``Open Data Model'' from 1998 \cite{Myers1998a}, the ``Semantic Web'' from 2001 \cite{BernersLee2001}, and ``A World Without Apps'' from 2019 \cite{BeaudouinLafon2019}), and it is needed for many forms of end-user programming (EUP) and programming by example (PBE) \cite{Myers1993}. However, it has always proven difficult to provide due to business reasons (companies want to keep their own data proprietary) and architectural challenges (it is very difficult to provide sufficient access without being overwhelming).

However, if an intelligent agent is going to be built at all, it will need significant access to this kind of data, so the privacy, security and architectural challenges will need to be addressed anyway.

\hypertarget{conclusions}{%
\section{Conclusions}\label{conclusions}}

Providing interactions in low-level toolkits and frameworks has always been the best way to make it easy (and even possible) for developers to include them in their applications. This will also be true for intelligent interaction techniques. Users will then have access to much better usability, effectiveness and accessibility within and across their applications. Another advantage of providing intelligence at the interaction technique level is that it will then behave \emph{consistently} across applications and be ubiquitously available by default. However, significant challenges remain, in all of the design, architecture, security, privacy, economic and business aspects.

\section*{Acknowledgements}

Some of the ideas in this paper were contributed by Motahhare Eslami, Dan Saffer, and Haiyi Zhu. Thanks also for contributions to this paper from Scott Hudson, Niki Kittur, Nik Martelaro, Bernita Myers, and John Zimmerman.

\renewcommand{\refname}{References}
\bibliographystyle{plainnat}
\bibliography{references}

\end{document}